\documentclass[twocolumn,showpacs,preprintnumbers,amsmath,amssymb]{revtex4}

\usepackage{graphicx}
\usepackage{dcolumn}
\usepackage{bm}

\begin{document}

\title{Schwinger Mechanism and Hawking Radiation as Quantum
Tunneling}
\author{Sang Pyo Kim}\email{sangkim@kunsan.ac.kr}
\affiliation{Department of Physics, Kunsan National University,
Kunsan 573-701, Korea} \affiliation{Asia
Pacific Center for Theoretical Physics, Pohang 790-784, Korea}

\date{}

\begin{abstract}
The common interpretation of the Hawking radiation as quantum
tunneling has some ambiguity such as coordinate-dependence of
tunneling rate and non-invariance of the action under canonical
transformations. It is shown that the tunneling process of black
holes can be successfully described by Rindler coordinates in
analogy with the Schwinger mechanism for pair production. We study
the tunneling process of a charged black hole and a BTZ black hole.
\end{abstract}

\maketitle

\section{Introduction}

More than a quarter century ago Hawking discovered that black holes
could radiate thermal radiations with the Hawking temperature
determined by the surface gravity at the event horizon
\cite{Hawking}. The surface gravity is the acceleration of a static
particle just outside the event horizon as measured at the spatially
infinity. Also Unruh observed that a uniformly accelerated particle
would experience a thermal state from the Minkowski vacuum with the
temperature determined by the acceleration \cite{Unruh}. Since then
there have been many attempts to derive or reinterpret the Hawking
radiation.

Recently the study of the Hawking radiation has been revived partly
because the Hawking radiation could be derived from the motion of
spherically emitted radiation including the back reaction of
radiation \cite{Kraus} and partly because quantum tunneling of
fluctuations or waves across the event horizon could lead to the
Hawking radiation \cite{Parikh}. In both cases the imaginary part of
action was responsible for tunneling of Hawking radiation.

Hawking radiation was originally derived by calculating the
amplitude ratio of an outgoing wave to an incoming wave scattered by
the event horizon. The recent quantum tunneling interpretation
differs from the early interpretation in that quantum fluctuations
just inside the horizon can cross the horizon and the tunneling
amplitude determines the Boltzmann distribution for Hawking
radiation
\cite{Srinivasan,Hemming,Vagenas,Medved,Shankaranarayanan,Angheben,Arzano,Zhang,Nadalini,Jiang,Kerner,Akhmedov,Wu,Gao}
(for review, see Ref. \cite{Padmanabhan}). However, there is some
ambiguity in the tunneling interpretation of Hawking radiation
\cite{Akhmedov}. The Boltzmann distribution depends on the
coordinates used to express the field equation. For instance, the
Schwarzschild coordinate leads to a temperature twice of the Hawking
temperature.

In this paper we show that the Rindler coordinates correctly provide
the tunneling rate for Hawking radiation and resolve the ambiguity.
We then apply the method to a charged black hole and a BTZ black
hole. The idea is based on the observation that the field equation
in the Rindler coordinates of a non-extremal black hole takes a
similar form as the field equation in an external electric field for
the Schwinger mechanism \cite{Schwinger}. In the space-dependent
(Coulomb) gauge the field equation conceptually leads to the
tunneling problem, in which virtual pairs of charged particles in
the Dirac sea experience a potential barrier lowered by the electric
field \cite{CNN,BPS,BMPS,Kim-Page,Kim-Page07,Dunne-Schubert}.
Further the tunneling rate is given by an instanton action in the
complex plane \cite{Kim-Page,Kim-Page07}. The tunneling rate
expressed by a contour integral in the complex plane is invariant
under canonical transformations.

The organization of this paper is as follows. In Sec. II, we briefly
review the Schwinger mechanism and introduce the tunneling rate as a
contour in the complex plane. In Sec. III, the tunneling rate for
Hawking radiation is formulated as a contour integral in Rindler
coordinates of black holes. We then apply the formula to a charged
black hole and a BTZ black hole.

\section{Schwinger Mechanism and Instanton Actions}

Schwinger used the proper-time method to calculate the effective
action of a charged particle in an external electromagnetic field
and found an imaginary part for a uniform electric field
\cite{Schwinger}. This leads to the vacuum decay through pair
production. The Schwinger mechanism for pair production by an
external field can be physically understood as follows. Virtual
pairs of charged particles from vacuum fluctuations can be separated
to become real pairs when the potential energy over the Compton
wavelength is greater than or equal to the rest mass energy. In the
tunneling interpretation the virtual pairs pass through the
potential barrier lowered by the electric field
\cite{CNN,BPS,BMPS,Kim-Page,Kim-Page07,Dunne-Schubert}. Roughly
speaking, the barrier width is inversely proportional to the
strength of the electric field.

The similarity between the Schwinger mechanism and the Hawking
radiation has been studied for many years
\cite{BPS,BMPS,Srinivasan}. A naive interpretation would be
separation of virtual pairs by an electric field in the former and
by the event horizon in the latter. Another interpretation would be
that virtual pairs accelerate by the electric field and these pairs
would feel thermal state from the Minkowski vacuum. Likewise the
Hawking radiation may be interpreted as the Unruh effect of a static
particle just outside the horizon with the acceleration of the
surface gravity measured at the infinity \cite{Unruh}. In fact, the
Rindler spacetime of the accelerated particle also has the horizons.

Now we further elaborate the analogy between the Schwinger mechanism
and the Hawking radiation. A difference from other studies is that
the Hawking radiation is interpreted as quantum tunneling in the
same way as the Schwinger mechanism. We observe that, for instance,
the Klein-Gordon equation for charged boson pairs in the
space-dependent (Coulomb) gauge takes the same form as the field
equation in the Rindler coordinates of black holes. Each Fourier
mode becomes a tunneling problem, the sub-barrier penetration. The
essence of the instanton action method is that the tunneling states
of this equation lead to the vacuum decay and pair production of
charged particles \cite{Kim-Page,Kim-Page07}. A similar form of
tunneling rate is provided by the worldline instanton
\cite{Dunne-Schubert}.

A constant electric field along the $x$-direction has the
space-dependent gauge potential $A_{\mu} = (-Ex, 0)$. The Fourier
mode of the Klein-Gordon equation for charge $e$ ($e > 0$) and mass
$\mu$ takes the form [in units with $\hbar = c = 1$ and with metric
signature $(-, +)$]
\begin{eqnarray}
\Bigl[ - \frac{\partial^2}{\partial x^2} - q(x) \Bigr]
\varphi_{\omega} (x) = 0, \quad q(x) = (\omega + eEx)^2 - \mu^2.
\label{tun eq}
\end{eqnarray}
In quantum mechanics, Eq. (\ref{tun eq}) is a tunneling problem with
the inverted potential barrier $- q(x)$ and $q(x)$ corresponds to
$p_x^2$ in the WKB approximation. Using the phase-integral formula
\cite{Froman}, the wave function can be written as
\begin{eqnarray}
\varphi_{\omega } (x) = A f_{\omega}+  B f_{\omega}^*,
\end{eqnarray}
where $f_{\omega} (x)$ is an asymptotic solution with unit incoming
flux and
\begin{equation}
A = (e^{2 S_{\omega}} +1 )^{1/2}, \quad B = e^{S_{\omega}}.
\end{equation}
Here the action can also be defined in the complex $x$-plane by the
contour integral \cite{Kim-Page07}
\begin{eqnarray}
{\cal S}_{\omega} = - i \oint_{\cal C} \sqrt{q(x)} dx = \frac{\pi
m^2 }{eE}, \label{cont int}
\end{eqnarray}
where the integral is along the contour in  Fig. 1.

\begin{figure}[t]
{\includegraphics[width=0.8\linewidth,height=0.04\textheight
]{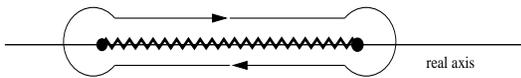}} \caption{The contour ${\cal C}$ in the complex
$x$-plane excluding a branch cut connecting two roots for
tunneling.}
\end{figure}

\section{Quantum Field Theory for Hawking Radiation}

Now we turn to quantum tunneling of Hawking radiation. Let us first
consider a charged black hole with mass $m$ and charge $q$
\begin{eqnarray}
ds^2 = - f (r)dt^2 + \frac{dr^2}{f(r)}
+ r^2 d \Omega_2^2,
\end{eqnarray}
where
\begin{eqnarray}
f (r) = 1 - \frac{2m}{r} + \frac{q^2}{r^2} = \frac{(r - r_+)(r - r_-)}{r^2}.
\end{eqnarray}
Here $r_{\pm} = m \pm \sqrt{m^2 - q^2}$ is the outer (inner) horizon.
The surface gravity is
\begin{eqnarray}
\kappa = \frac{f'(r_+)}{2} = \frac{r_+ - r_-}{2 r_+^2},
\end{eqnarray}
from which follows the Hawking temperature
\begin{eqnarray}
T_H = \frac{\kappa}{2 \pi}. \label{hawking}
\end{eqnarray}

The $s$-wave of a scalar field $\Phi$ with mass $\mu$ satisfies
\begin{eqnarray}
\Bigl[- \frac{1}{f} \frac{\partial^2 }{\partial t^2} + \frac{1}{r^2}
\frac{\partial}{\partial r} \Bigl( f r^2 \frac{\partial }{\partial r} \Bigr)
- \mu^2 \Bigr] \Phi = 0.
\end{eqnarray}
The solution of the form $\Phi = e^{- i \omega t + i S(r)}$ may
describe quantum tunneling of Hawking radiation when the action $S$
has an imaginary part and thus the amplitude squared $|\Phi|^2 =
e^{- 2 S_{I}}$ shows tunneling across the horizon. The action in the
WKB approximation
\begin{eqnarray}
S (r) = \pm \int_{r_0}^r \Bigl( \omega^2
- m^2 f \Bigr)^{1/2} \frac{dr}{f} \label{act}
\end{eqnarray}
with $r_0$ inside the horizon has a simple pole at the horizon $r_+$.
The positive (negative) sign is for an outgoing (incoming) wave.

The imaginary part obtained by taking a semi-circle over the horizon
$r_+ $ is
\begin{eqnarray}
S_{I} = \pi \omega \Bigl( \frac{r_+^2}{r_+ - r_-} \Bigr) =
\frac{\pi \omega}{2 \kappa}.
\end{eqnarray}
Then the tunneling rate is given by
\begin{eqnarray}
P (\omega) = e^{- \frac{\pi \omega}{\kappa} } = e^{- \omega /T}.
\end{eqnarray}
Here the temperature $T = \kappa /\pi$ is the twice of the Hawking
temperature (\ref{hawking}). This ambiguity is closely related with
coordinates used to calculate the tunneling rate. Several ways have
been proposed so far to remedy this discrepancy such as the ratio of
emission to absorption \cite{Srinivasan} and a proper distance along
the radial direction \cite{Angheben}. Another ambiguity is that the
action (\ref{act}) and the imaginary part is not invariant under
canonical transformations \cite{Akhmedov}.

We calculate the tunneling rate of $s$-waves (two-dimensional
sector) in Rindler coordinates of a non-extremal charged black hole
and a BTZ black hole. Introducing $r^* = \int dr/ f(r)$, we write
the metric in a conformal spacetime
\begin{eqnarray}
ds^2 = - f(r^*) (dt^2 - dr^{*2}),
\end{eqnarray}
where the Regge-Wheeler coordinate for a non-extremal black hole
$(m^2 > q^2)$ is
\begin{eqnarray}
r^* = r + \frac{r_+^2}{r_+ - r_-} \ln (r - r_+) - \frac{r_-^2}{r_+ - r_-}
\ln (r - r_-).
\end{eqnarray}
Then the spatial part of the scalar, $\Phi = e^{- i \omega t}
\varphi_{\omega} (r^*)$, takes the form
\begin{eqnarray}
\Bigl[- \frac{\partial^2}{\partial r^{*2}}  - q_{\omega} (r^*)
\Bigr] \varphi_{\omega} = 0, \quad q_{\omega} (r^*) = \omega^2 -
\mu^2 f(r^*). \label{bh mode}
\end{eqnarray}
The waves (particles) can be observed at spatial infinity only for
$\omega \geq m$. From the analogy with the Schwinger mechanism, the
tunneling rate (imaginary part of the action) is determined by the
contour integral in the complex $r$-plane
\begin{eqnarray}
{\cal S}_{\omega} =  i \oint_{\cal C} \sqrt{q_{\omega}(r^*)} dr^*.
\label{bh con}
\end{eqnarray}
Here the contour integral is taken the inside of ${\cal C}$ in
contrast with the outside of ${\cal C}$ in Sec. II.

However, as the simple pole occurs at $r^* = - \infty$, the contour
integral (\ref{bh con}) may not be appropriate for practical
calculations. Instead, in the Rindler coordinate
\begin{eqnarray}
f (r) = \frac{r_+^2}{r^2} (\kappa x)^2,
\end{eqnarray}
the black hole spacetime becomes
\begin{eqnarray}
ds^2 &=& - \frac{(\kappa r_+)^2}{(m + \sqrt{\bar{m}^2
+ (\kappa r_+ x)^2})^2} (x dt)^2 \nonumber\\
&&+ \frac{(\kappa r_+)^2 (m + \sqrt{\bar{m}^2 +  (\kappa r_+
x)^2})^2}{\bar{m}^2 + (\kappa r_+ x)^2} dx^2, \label{rn bh}
\end{eqnarray}
where $\bar{m} = (r_+ - r_-)/2 = \sqrt{m^2 - q^2}$. Near the event
horizon the spacetime (\ref{rn bh}) approximately takes the form of
$ds^2 = - (\kappa x)^2 dt^2 + dx^2$, a Rindler spacetime with
acceleration $\kappa$. We extend the right wedge of the Minkowski
spacetime (\ref{rn bh}) exterior to the horizon to the left wedge,
which is an analytical continuation of the interior to the horizon.
The simple pole is located at $x = 0$. Now the action for tunneling
\begin{eqnarray}
{\cal S}_{\omega} =  i \oint_{\cal C} \sqrt{q_{\omega}(x)}
\frac{dx}{(\kappa x)} = \frac{2 \pi \omega}{\kappa},
\end{eqnarray}
leads to the tunneling rate
\begin{eqnarray}
P (\omega) = e^{- {\cal S}_{\omega}} = e^{- \omega /T_H}.
\label{tunneling rate}
\end{eqnarray}

We compare with other coordinates. In Ref. \cite{Angheben} the
proper distance is used as a coordinate to calculate the tunneling
rate. The proper distance, $\sigma = \int dr / \sqrt{f}$, measured
from the event horizon is
\begin{eqnarray}
\sigma &=& \sqrt{(r - r_+)(r - r_-)} \nonumber\\
&& + m \ln \Bigl( \frac{r -  m + \sqrt{(r - r_+)(r - r_-)}}{m}
\Bigr).
\end{eqnarray}
Then the two-dimensional metric near the event horizon
\begin{eqnarray}
ds^2 = - (\kappa \sigma)^2 dt^2 + d \sigma^2,
\end{eqnarray}
is the Rindler spacetime. This is the reason why the proper distance
method recovers the correct tunneling rate. On the other hand, the
isotropic coordinate defined as
\begin{eqnarray}
\int \frac{d \rho}{\rho} = \int \frac{dr}{r \sqrt{f(r)}},
\end{eqnarray}
is
\begin{eqnarray}
r = \frac{(\bar{\rho} + \bar{m} )^2 - q^2}{ 2 (\bar{\rho} - r_- )},
\end{eqnarray}
where $\bar{\rho} = \rho / c$ and $c$ is a constant. Later we choose
$c = r_+/m$  to match the isotropic spacetime near the horizon with
the Rindler spacetime. The metric is
\begin{eqnarray}
ds^2 &=& - \frac{[(\bar{\rho} + \bar{m})^2 - q^2
- 2 r_+ (\bar{\rho} - r_-)]}{[(\bar{\rho} + \bar{m})^2 - q^2]} \nonumber\\
&& \times \frac{[(\bar{\rho} + \bar{m})^2 - q^2
- 2 r_- (\bar{\rho} - r_-)]}{[(\bar{\rho} + \bar{m})^2 - q^2]}
dt^2 \nonumber\\
&& + \frac{m^2 [(\bar{\rho} + \bar{m})^2 - q^2]^2}{
4 r_+^2\bar{\rho}^2 (\bar{\rho} - r_-)^2} d \rho^2.
\end{eqnarray}
Near the horizon it becomes again the Rindler spacetime witt the
acceleration $\kappa$. Thus the isotropic coordinates also recover
the correct result.

Finally, we consider the BTZ black hole with the metric
\begin{eqnarray}
ds^2 = - g (r) dt^2 + \frac{dr^2}{g(r)} + r^2 \Bigl(
d \theta - \frac{j}{2r^2} dt \Bigr)^2,
\end{eqnarray}
where
\begin{eqnarray}
g (r) = - m +  \frac{j^2}{4 r^2} + \lambda r^2.
\end{eqnarray}
The BTZ black hole has the outer (inner) horizon at $r_{\pm}^2 = (m
\pm \sqrt{m^2 - \lambda j^2})/(2 \lambda)$. The angular part $d \chi
= d \theta - j dt/(2r^2)$ describes a rotation. For the $s$-wave the
angular part does not play any role. The remaining two-dimensional
sector can be treated in a similar way. Introducing the coordinate
\begin{eqnarray}
g = \frac{\lambda (r^2 - r_+^2) (r^2 - r_-^2)}{2 r^2} = (\kappa x)^2,
\end{eqnarray}
we find the metric
\begin{eqnarray}
ds^2 &=& - (\kappa x)^2 dt^2 + \frac{8 \kappa^2}{\lambda}\nonumber\\
&\times& \frac{\Bigl(1 + \frac{m + 4 (\kappa x)^2}{\sqrt{(m + 4 (\kappa x)^2)^2 -
\lambda j^2}} \Bigr)^2}{m +
4 (\kappa x)^2 + \sqrt{(m + 4 (\kappa x)^2)^2 - \lambda j^2}} dx^2.
\label{btz rin}
\end{eqnarray}
When $\kappa$ is the surface gravity
\begin{eqnarray}
\kappa = \frac{\lambda}{4} \frac{r_+^2 - r_-^2}{r_+} = \frac{\sqrt{m^2 -
\lambda j^2}}{4 r_+},
\end{eqnarray}
the spacetime (\ref{btz rin}) near the horizon becomes a Rindler
spacetime. Therefore the tunneling rate of  BTZ black hole is also
given by (\ref{tunneling rate}).

\section{Conclusion}

We critically reviewed the current study of the Hawking radiation as
quantum tunneling. The common method to calculate the tunneling rate
of a field across the event horizon has been confronted with two
ambiguities: the coordinate-dependence of the action and the
temperature different from the Hawking temperature \cite{Akhmedov}.
Though some prescriptions have been put forth such as the proper
distance and the ratio of emission to absorption, the proper choice
of coordinate has been an issue of debate.

In this paper we used the Rindler coordinates as an appropriate
spacetime to describe the tunneling problem of waves across the
event horizon in the black hole spacetime. The reasons are that
tunneling occurs across the event horizon and the tunneling rate is
determined by the geometric property at the horizon while the
Rindler spacetime approximately describes the spacetime near the
horizon. Further the Hawking radiation can be interpreted as the
Unruh effect of an observer accelerating with the surface gravity.
Another interesting observation is that the field equation in a
Rindler spacetime is reminiscent of the field equation for the
Schwinger mechanism. The rate of pair production is obtained by a
contour integral of the action in the complex plane and similarly
the tunneling rate for black hole is given by the contour integral
in the complex plane of the Rindler coordinate.

We applied the tunneling rate in the form of contour integral to a
charged black hole and a BTZ black hole. In both cases the black
hole has the Rindler coordinates in an explicit form except for the
extremal black hole. Our tunneling formula leads to the correct
Boltzmann distribution and the Hawking temperature. Further this
formula avoids the ambiguity of tunneling interpretation. A caveat
is that the formula cannot be applied to the extremal black hole
because it does not have a Rindler spacetime near the horizon. The
Hawking temperature vanishes there and thus no Hawking radiation.
The development of field theoretical interpretation of tunneling in
Rindler coordinates and applications to other black holes will be
addressed in a future publication \cite{Kim07}.

\acknowledgments The author would like to thank participants of New
Frontiers of Black Hole Physics at APCTP for useful comments and
discussions. This work was supported by the Korea Science and
Engineering Foundation (KOSEF) grant funded by the Korea government
(MOST) (No. F01-2007-000-10188-0).


\begin{thebibliography}{99}

\bibitem{Hawking} S.~W.~Hawking, Comm. Math. Phys. {\bf 43}, 199
(1975).

\bibitem{Unruh} W.~G.~Unruh, Phys. Rev. D {\bf 14}, 870 (1976).

\bibitem{Kraus} P.~Kraus and F.~Wilczek, Nucl. Phys. B {\bf 433},
403 (1995); {\it ibid.} {\bf 437}, 231 (1995); P.~Kraus and
E.~Keski-Vakkuri, {\it ibid.} {\bf 491}, 249 (1997).

\bibitem{Parikh} M.~K.~Parikh and F.~Wilczek, Phys. Rev. Lett.
{\bf 85}, 5042 (2000).

\bibitem{Srinivasan} K.~Srinivasan and T.~Padmanabhan, Phys. Rev. D {\bf 60},
024007 (1999).

\bibitem{Hemming} S.~Hemming and E.~Keski-Vakkuri, Phys. Rev. D {\bf
64}, 044006 (2001).

\bibitem{Vagenas} E.~C.~Vagenas, Phys. Lett. B {\bf 503}, 399
(2001); {\it ibid.} {\bf 533}, 302 (2002).

\bibitem{Medved} A.~J.~M.~Medved, Phys. Rev. D {\bf 66}, 124009
(2002).

\bibitem{Shankaranarayanan} S.~Shankaranarayanan, T.~Padmanbhan, and
K.~Srinivasan, Class. Quantum Grav. {\bf 19}, 2671 (2002);
S.~Shankaranarayanan, Phys. Rev. D {\bf 67}, 084026 (2003).

\bibitem{Angheben} M.~Angheben, M.~Nadalini, L.~Vanzo, and S.~Zerbini,
JHEP {\bf 0505}, 014 (2005).

\bibitem{Arzano} M.~Arzano, A.~J.~M.~Medved, and E.~C.~Vagenas,
JHEP {\bf 0509}, 037 (2005).

\bibitem{Zhang} J.~Zhang and Z.~Zhao, JHEP {\bf 0510}, 055 (2005);
 Nucl. Phys. B {\bf 725}, 173 (2005); Phys. Lett. B {\bf 638}, 110
(2006).

\bibitem{Nadalini} M.~Nadalini, L.~Vanzo, and S.~Zerbini, J. Phys. A:
Math. Gen. {\bf 39}, 6601 (2006).

\bibitem{Jiang} Q.-Q.~Jiang, S.-Q.Wu, and X.~Cai, Phys. Rev. D {\bf
73}, 064003 (2006).

\bibitem{Kerner} R.~Kerner and R.~B.~Mann, Phys. Rev. D {\bf 73},
104010 (2006).

\bibitem{Akhmedov} E.~T~Akhmedov, V.~Akhmedova, and D.~Singleton,
Phys. Lett. B {\bf 642}, 124 (2006).

\bibitem{Wu} X.~Wu and S.~Gao, Phys. Rev. D {\bf 75}, 044027 (2007).

\bibitem{Gao} L.~Gao, H.~Zhang, and W.~Liu, Int. J. Theor. Phys.
{\bf 46}, 33 (2007).

\bibitem{Padmanabhan} T.~Padmanabhan, Phys. Rep. {\bf 406}, 49 (2005).

\bibitem{Schwinger} J.~Schwinger, Phys. Rev. {\bf 82}, 664 (1951).

\bibitem{CNN} A.~Casher, H.~Neuberger, and S.~Nussinov, Phys. Rev. D {\bf 20},
179 (1979); H.~Neuberger, {\it ibid.} {\bf 20}, 2936 (1979).

\bibitem{BPS} R.~Brout, R.~Parentani, and Ph.~Spindel, Ann. Phys.
{\bf 353}, 209 (1991); R.~Parentani and R.~Brout, {\it ibid.} {\bf
388}, 474 (1992); R.~Parentani and S.~Massar, Phys. Rev. D {\bf 55},
3603 (1997).

\bibitem{BMPS} R.~Brout, S.~Massar, R.~Parentani, and Ph.~Spindel,
Phys. Rep. {\bf 260}, 329 (1995).

\bibitem{Kim-Page} S.~P.~Kim and D.~N.~Page, Phys. Rev. D {\bf 65}, 105002
(2002); {\it ibid.} {\bf 73}, 065020 (2006).

\bibitem{Kim-Page07} S.~P.~Kim and D.~N.~Page, Phys. Rev. D {\bf 75}, 045013 (2007).

\bibitem{Dunne-Schubert} G.~V.~Dunne and C.~Schubert, Phys. Rev. D {\bf 72},
105004 (2005); G.~V.~Dunne, Q.-H.~Wang, H.~Gies, and C.~Schubert,
{\it ibid.} {\bf 73}, 065028 (2006).

\bibitem{Froman} N.~Fr\"{o}man and P.~O.~Fr\"{o}man, Nucl. Phys.
{\bf A147}, 606 (1970); {\it Phase-Integral Method}
(Springer-Verlag, New York, 1996).

\bibitem{Kim07} S.~P.~Kim, ``Hawking Radiation as Quantum Tunneling in
Rindler Coordinates,'' in prepration.

\end{thebibliography}
\end{document}